\documentclass[12pt]{scrartcl}
\usepackage{geometry}
\usepackage{mathtools}
\usepackage[boxruled,linesnumbered]{algorithm2e}
\usepackage{blindtext, graphicx}
\usepackage{url}
\usepackage{relsize}
\usepackage{cite}
\usepackage[breaklinks,colorlinks=true,citecolor=blue,pdftex]{hyperref}
\usepackage{bm}
\usepackage{changepage}

\begin{document}

\title{Comparison of Flow Scheduling Policies for Mix of Regular and Deadline Traffic in Datacenter Environments}

\subtitle{\lbrack Technical Report\rbrack}

\author{\large
    Mohammad Noormohammadpour and Cauligi S. Raghavendra\\ 
    \textit{\large Ming Hsieh Department of Electrical Engineering}\\ 
    \textit{\large University of Southern California (USC)}
}

\date{}

\maketitle

\begin{abstract}
Datacenters are the main infrastructure on top of which cloud computing services are offered. Such infrastructure may be shared by a large number of tenants and applications generating a spectrum of datacenter traffic. Delay sensitive applications and applications with specific Service Level Agreements (SLAs), generate deadline constrained flows, while other applications initiate flows that are desired to be delivered as early as possible. As a result, datacenter traffic is a mix of two types of flows: deadline and regular. There are several scheduling policies for either traffic type with focus on minimizing completion times or deadline miss rate. In this report, we apply several scheduling policies to mix traffic scenario while varying the ratio of regular to deadline traffic. We consider FCFS (First Come First Serve), SRPT (Shortest Remaining Processing Time) and Fair Sharing as deadline agnostic approaches and a combination of Earliest Deadline First (EDF) with either FCFS or SRPT as deadline-aware schemes. In addition, for the latter, we consider both cases of prioritizing deadline traffic (Deadline First) and prioritizing regular traffic (Deadline Last). We study both light-tailed and heavy-tailed flow size distributions and measure mean, median and tail flow completion times (FCT) for regular flows along with Deadline Miss Rate (DMR) and average lateness for deadline flows. We also consider two operation regimes of lightly-loaded (low utilization) and heavily-loaded (high utilization). We find that performance of deadline-aware schemes is highly dependent on fraction of deadline traffic. With light-tailed flow sizes, we find that FCFS performs better in terms of tail times and average lateness while SRPT performs better in average times and deadline miss rate. For heavy-tailed flow sizes, except for tail times, SRPT performs better in all other metrics.
\end{abstract}

\section{Introduction}
Cloud computing provides an efficient computing platform for many applications. Some benefits include easy scaling of applications as well as reduced maintenance and management costs by allowing some level of statistical multiplexing. These services are offered on top of a datacenter infrastructure. Cloud datacenters often host a wide variety of applications that generate traffic flows with different requirements. Datacenter flows can generally be classified as deadline and non-deadline (regular) flows \cite{karuna}. 

\textbf{Deadline flows:} Latency sensitive applications require timely completion of traffic flows. In many cases, completing flows past the deadlines is useless \cite{pdq}. A major application that generates such flows is search. In such applications, there are multiple aggregation levels where query results are combined and then passed to the upper level \cite{dctcp, ictcp}. If an aggregator does not receive a query's reply by the deadline, that query will be left out \cite{bing}. For these flows, \textit{deadline miss rate} is a useful performance metric since it determines the fraction of queries left out. For some other applications, delivery after the deadline may still be valuable or even necessary. For instance, important periodic backups, search index synchronization and content replication are operations that need be completed even if they miss their deadlines \cite{calendaring, swan, b4}. For these flows, \textit{average lateness} is an applicable performance metric since it shows how much later than expected flows complete.

\textbf{Regular (non-deadline) flows:} Many applications generate traffic flows that do not have deadlines. It is however desired that flows complete as early as possible since they play a role in overall running time of such applications. For example, background operations such as VM migration, data warehousing as well as some long running batch computing and data analytics jobs may generate such flows. 

Effective mix traffic scheduling requires attention to objectives of both regular and deadline traffic. It is desired that regular flows finish as early as possible while deadline flows are required to complete prior to their deadlines. A prior work models this as stochastic optimization problem with the objective of minimizing per-packet latency \cite{karuna}. Prior work however does not study the variety of well-known scheduling policies for mix traffic scenario.

In this report, we focus on abstract evaluations and comparison of scheduling policies using artificially generated traffic that follows specific distributions (Exponential and Pareto). We perform simulations to explore effectiveness of a range of scheduling policies for mix flow scheduling. We consider mix traffic with various demand ratios between deadline and regular flows starting from mostly deadline traffic and ending with mostly regular traffic. We experiment with both light-tailed and heavy-tailed demand distribution scenarios along with two operation regimes of lightly-loaded and heavily-loaded. 

We find that performance of deadline aware schemes (combination of EDF with SRPT and FCFS) is highly dependant on fraction of deadline traffic. With light-tailed flow sizes, we find that FCFS performs better in terms of tail times and average lateness while SRPT performs better in average times and deadline miss rate. For heavy-tailed flow sizes, except for tail times, SRPT performs better in all other metrics.

\section{Scheduling Policies}
A variety of flow scheduling techniques have been applied in datacenters at the data plane layer. In-network scheduling is usually enforced at the outgoing ports of network switches where flows going over same links share buffer space. The scheduling policy determines how these flows interact with each other and will affect the completion times of regular flows as well as deadline miss rate for deadline flows.

A wide range prior work is based on switches applying FCFS policy \cite{dctcp, ictcp, otcp, cut-payload} considering the lower cost and wider use of switches with a single outgoing queue per port. If multiple queues are available per outgoing port with the ability to assign priorities to such queues, it is possible to approximate the SRPT policy by assigning smaller flows to higher priority queues \cite{pias, pfabric, pase, detail}. Many recent switches offer up to $8$ priority queues per outgoing port \cite{pfabric, detail}. Priorities can be assigned to packets either at the end-hosts \cite{pias} or using dynamic packet counting techniques at the switches \cite{dpp}. Solutions have also been proposed to enforce EDF policy using both switches and end-points as part of scheduling process \cite{pdq}.

Trade-offs are involved in application of different scheduling policies. FCFS has been shown to provide bounded lateness \cite{FCFS-tardiness} but can lead to increased deadline misses or completion times depending on arrival orders. Fair Sharing prevents the starvation problem for large flows, however it is not effective in minimizing completion times \cite{srpt_fairness} nor number of missed deadlines \cite{edf}. SRPT is the optimal policy in minimizing mean flow completion times. It however may lead to increased tail completion times and starvation of some flows. EDF is the optimal scheduling policy for deadline traffic in case all deadlines can be met. However, its performance degrades as the number of deadlines that cannot be met increases. In contrast, SRPT has been shown to perform best in meeting deadlines under heavy loads although it performs poorly under light loads \cite{edf-vs-srpt}.

\begin{table}[t!]
\centering
\begin{tabular}{|p{3cm}|p{9cm}|p{3.3cm}|} 
    \hline
    Policy & Description & Is Preemptive? \\
    \hline
    \hline
    FCFS  & Requests are scheduled according to their arrival time. Deadline and non-deadline traffic are treated the same way.  & No \\
    \hline
    SRPT  & Requests are scheduled according to their remaining processing time. If there is no new arrival, the request scheduled at current timeslot will also be scheduled at the next timeslot since it will still be the job with minimal remaining demand. Deadline and non-deadline traffic are treated the same way.  & Yes \\
    \hline
    Fair Sharing  & Applies max-min fairness \cite{mmf} to all requests. Each request will receive at most as much as its fair share. Deadline and non-deadline traffic are treated the same way.  & Yes \\
    \hline
    EDF-FCFS-DF (Deadline First)  & Combines EDF and FCFS applying former to deadline traffic and latter to non-deadline traffic. Deadline traffic is strictly prioritized over non-deadline traffic.  & Yes \\
    \hline
    EDF-SRPT-DF (Deadline First)  & Combines EDF and SRPT applying former to deadline traffic and latter to non-deadline traffic. Deadline traffic is strictly prioritized over non-deadline traffic.  & Yes \\
    \hline
    EDF-FCFS-DL (Deadline Last)  & Combines EDF and FCFS applying former to deadline traffic and latter to non-deadline traffic. Non-deadline traffic is strictly prioritized over deadline traffic.  & Non-deadline: No ~~~~~~ Deadline: Yes  \\
    \hline
    EDF-SRPT-DL (Deadline Last)  & Combines EDF and SRPT applying former to deadline traffic and latter to non-deadline traffic. Non-deadline traffic is strictly prioritized over deadline traffic.  & Yes \\
    \hline
\end{tabular}
\caption{Scheduling policies considered in this report} \label{table_policies}
\end{table}

\section{Problem Setup}
We consider an online scenario where flows may arrive anytime with no prior knowledge of their arrival. We assume all traffic is going over a single path. We also assume that all arriving traffic is admitted and scheduled even in case deadlines cannot be met. In general it may be possible to verify that deadlines can be met upon arrival of flows \cite{amoeba, rcd, dcroute}. Study of how admission control affects performance is out of the scope of this report.

\textbf{Traffic:} The traffic consists of a set of flows $F_{i},~i \in \{1 \dots N\}$. Datacenter traffic is a mix of two types of flows: non-deadline (regular) and deadline. Regular flows can be shown with two tuples $(A_{i},V_{i})$ which determine their arrival times and volumes (also known as demand, size). Deadline flows are each shown with three tuples $(A_{i},V_{i},D_{i})$ representing flows' arrival times, volumes and deadlines. In general, volumes are either known or can be estimated in most cases \cite{d3}. In our experiments, we assumed flows with known volumes.

\textbf{Distribution Patterns:} We assumed a Poisson distribution for the arrival times and changed the arrival rate of flows to vary the offered load. In general, datacenter flow arrivals may follow a short-lived bursty pattern \cite{wild, nature, traffic_dc_char, social_inside}. We aim to study bursty arrivals in the future using real traces of traffic.

We considered two distribution patterns for flow sizes: light-tailed and heavy-tailed. Prior measurements on datacenter traffic have shown more similarity to heavy-tailed distributions with majority of bytes being moved by a minority of large flows \cite{social_inside, wild, dctcp, vl2}. To examine light-tailed, we used the Exponential distribution and for heavy-tailed, we applied Pareto distribution setting the scale and shape parameters so that both distributions have the same mean.

Finally, datacenter traffic carries two types of deadline traffic. Flows with short deadlines that represent applications such as search, online ads and recommendation as well as flows with long deadlines for backup, synchronization and content replication. As a result, we classified large deadline flows as flows with soft deadlines while short deadline flows as flows with hard deadlines.

\textbf{Regimes:} By changing the arrival rate of requests, we considered two operation regimes. In the lightly-loaded regime, the average utilization of the link is far from $100\%$ while in the heavily-loaded regime it is close to $100\%$. In general, there may be intervals of time when the network is overloaded, however the average utilization in datacenters is generally low \cite{wild, social_inside}.

\textbf{Performance Metrics:} We focus on Flow Completion Time (FCT) of regular flows as well as Deadline Miss Rate (DMR) and average lateness (i.e., how long subsequent to deadlines flows completed, in case deadlines were missed) for deadline flows. We considered three metrics of Average FCT (AFCT), Median FCT (MFCT) and Tail FCT (TFCT) for regular flows. DMR is calculated for all deadline flows, however average lateness is computed for only soft deadline flows.

\textbf{Mix Traffic Scheduling Policies:} We considered a combination of policies to handle mix traffic as shown in table \ref{table_policies}. In cases when we applied EDF, we needed a complementary policy to help us schedule the non-deadline traffic. We also studied both cases of strictly prioritizing deadline traffic (Deadline First) and non-deadline traffic (Deadline Last).

\begin{figure}
    \centering
    \includegraphics[width=\textwidth]{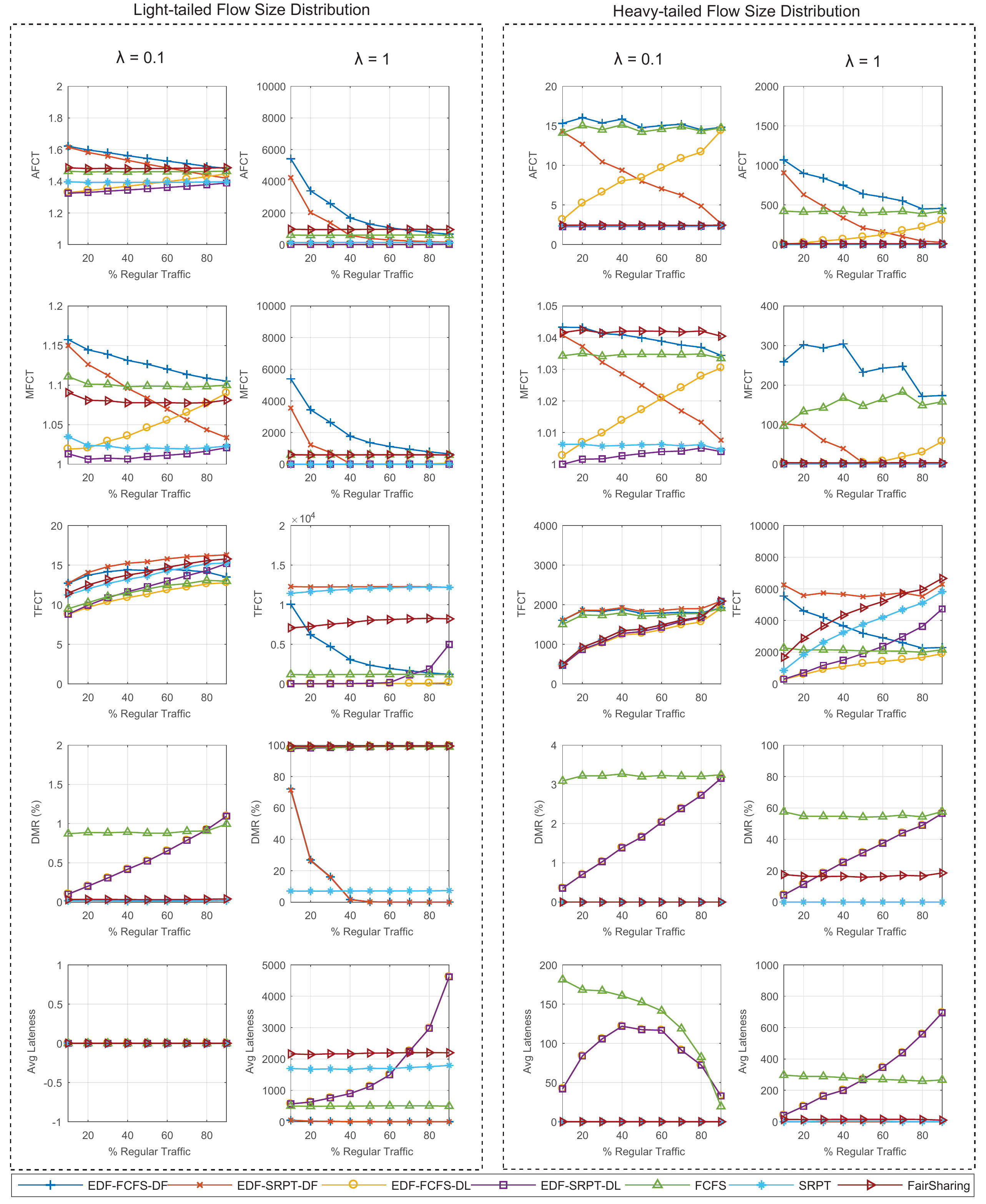}
    \caption{Performance of Traffic Scheduling Policies}
    \label{fig:dist}
\end{figure}

\section{Experiments}
In this section, we perform four experiments with two different operational regimes and flow size distributions. We first explain the simulation parameters and method and then move on to discussing the results.

\textbf{Simulation Method:} We considered small timeslots of length $\delta << 1$. The transmission rate of a flow is constant during each timeslot. Upon beginning of every slot, the rate is recalculated considering the new flow arrivals during the last timeslot. By making timeslots short enough, this turns into an almost real-time scenario.

\textbf{Simulation Parameters:} A path with capacity of $1$ was assumed. We considered two arrival rates of $\lambda=0.1$ (average utilization about $10\%$) and $\lambda=1$ (average utilization above $95\%$) for the Poisson distribution. For flow sizes, both Exponential and Pareto distributions were configured with an average of $1$ units of traffic. The minimum value for Pareto distribution was set to $0.1$ units. The simulation was performed until completion of $10000$ flow arrivals. We varied the volume of regular traffic from $10\%$ of total traffic units to $90\%$. Deadline requests with a size greater than $2$ times the mean were considered as flows with soft deadlines while other deadline requests were considered to have a hard deadline. We set $\delta = 0.1$ units of time. Reducing $\delta$ further did not have any noticeable effect on the results. We repeated all experiments $20$ times and calculated the average.

\subsection{Results}
Figure \ref{fig:dist} shows our simulation results. The top three charts in each column are normalized by the minimum value in those charts. We have reported how different disciplines perform regarding each one of the metrics in tables \ref{table:fct} through \ref{table:lateness}.

\begin{table}[t!]
\centering
\begin{tabular}{|p{3cm}|p{6.5cm}|p{6.5cm}|} 
    \cline{2-3}
    \multicolumn{1}{c|}{} & Light-tailed & Heavy-tailed \\
    \hline
    Lightly loaded (Utilization $\approx 10\%$) & EDF based schemes that prioritize regular traffic offer the minimal AFCT with small volume of regular traffic while converging to the complementary non-deadline policy (FCFS, SRPT) as volume of regular traffic increases. SRPT performs better than FCFS for average and median and worse for tail and also always better than Fair Sharing. FCFS works better than Fair Sharing for tail and average cases but worse for median. EDF based schemes that prioritize deadline traffic perform worse than others with small regular traffic but converge to the complementary non-deadline policy (FCFS, SRPT) as volume of regular traffic increases. & EDF based schemes that prioritize regular traffic offer the minimal AFCT with small volume of regular traffic while converging to the complementary non-deadline policy (FCFS, SRPT) as volume of regular traffic increases. SRPT performs better than FCFS and Fair Sharing in all cases. FCFS performs better than Fair Sharing for average and tail cases while Fair Sharing performs better for median. EDF based schemes that prioritize deadline traffic perform worse than others with small regular traffic but converge to the complementary non-deadline policy (FCFS, SRPT) as volume of regular traffic increases. \\
    \hline
    Heavily loaded (Utilization $> 90\%$) & Almost the same behavior as above. Fair Sharing and FCFS behave almost similarly in the median this time. SRPT performs far worse than FCFS and Fair Sharing in the tail. & Almost same as above. FCFS performs far worse compared to both SRPT and Fair Sharing in average and median. For the tail, FCFS performs much better when there is mostly regular traffic. Also, SRPT performs better than Fair Sharing for the tail. \\
    \hline
\end{tabular}
\caption{Performance analysis of scheduling policies for average, median and tail FCT (Only for regular traffic), rows $1-3$ in Figure \ref{fig:dist}} \label{table:fct}
\end{table}

\begin{table}[t!]
\centering
\begin{tabular}{|p{3cm}|p{6.5cm}|p{6.5cm}|} 
    \cline{2-3}
    \multicolumn{1}{c|}{} & Light-tailed & Heavy-tailed \\
    \hline
    Lightly loaded (Utilization $\approx 10\%$) & FCFS results in maximum miss rate. Also EDF schemes that prioritize regular traffic miss more deadlines when volume of regular traffic increases. & Same result as light-tailed. \\
    \hline
    Heavily loaded (Utilization $> 90\%$) & SRPT performs stably and keeps miss rate down for all ratios of regular to deadline traffic. FCFS and Fair Sharing miss almost all deadlines. EDF based schemes with deadline traffic prioritized miss a lot more deadlines when most traffic is deadline traffic. This drops as deadline traffic decreases to almost zero when half the traffic has deadlines. & Almost similar to above with Fair Sharing missing more deadlines but much less than FCFS. SRPT misses no deadlines. \\
    \hline
\end{tabular}
\caption{Performance analysis of scheduling policies for deadline miss rate (Only for deadline traffic), row $4$ in Figure \ref{fig:dist}} \label{table:dmr}
\end{table}

\begin{table}[t!]
\centering
\begin{tabular}{|p{3cm}|p{6.5cm}|p{6.5cm}|} 
    \cline{2-3}
    \multicolumn{1}{c|}{} & Light-tailed & Heavy-tailed \\
    \hline
    Lightly loaded (Utilization $\approx 10\%$) & All soft deadline traffic meet their deadlines. & Average lateness for Fair Sharing, SRPT and EDF based schemes that prioritize deadline traffic is almost zero. EDF based schemes that prioritize regular traffic perform worse when there is equal share of regular and deadline traffic. FCFS performs worst. \\
    \hline
    Heavily loaded (Utilization $> 90\%$) & EDF based schemes that prioritize deadline traffic perform best. FCFS performs better than SRPT and that better than Fair Sharing. EDF based schemes that prioritize regular traffic perform similar to FCFS when most traffic has deadlines but grow far worse than all schemes when most traffic is regular. & Almost similar to above with performance of FCFS being constant as ratio of deadline to regular traffic increases. EDF based schemes that prioritize regular traffic perform worse as we increase the regular traffic. \\
    \hline
\end{tabular}
\caption{Performance analysis of scheduling policies for average lateness (Only for soft deadline traffic), row $5$ in Figure \ref{fig:dist}} \label{table:lateness}
\end{table}

\section{Conclusions}
The variety of applications that run in datacenters creates a diverse set of traffic flows with different properties. Flows can be generally divided into deadline and regular flows depending on whether they have requirements on completion times. Also, deadline flows can generally be categorized into soft and hard deadline flows. For regular flows, flow completion time is the performance metric while deadline miss rate determines the network performance for deadline flows. For flows with soft deadlines, if deadlines cannot be met, it is still valuable to complete them in which case lateness is an important performance metric. In this report, we performed experiments to evaluate scheduling policies for mix traffic scenario. We considered lightly and heavily loaded regimes as well as light-tailed and heavy-tailed flow size distributions to test well-known scheduling policies. Our results show that SRPT is the best scheduling policy for heavy-tailed flow sizes in case reducing tail times is not an objective. SRPT also performs well for light-tailed distributions except for average lateness (in addition to tail times). We also find that EDF based policies do not perform well under the variety of loads and distributions and that their effectiveness is highly contingent upon ratio of regular to deadline traffic.

{\footnotesize \bibliographystyle{unsrt}
\bibliography{citations}}

\end{document}